\documentclass[prd,twocolumn,preprintnumbers,showpacs,amsmath,amssymb,nofootinbib]{revtex4}
\usepackage{latexsym,ifthen,graphics,color,epsfig,hyperref}

\newcommand{\ie}{{i.e.}}

\newcommand{\eg}{{e.g.}}
\newcommand{\aka}{{a.k.a.}}

\newcommand{\viz}{{viz.}}
\newcommand{\wrt}{with respect to}

\newcommand{\lhs}{left-hand side}

\newcommand{\rhs}{right-hand side}

\newcommand{\be}{\begin{equation}}
\newcommand{\ee}{\end{equation}}
\newcommand{\bea}{\begin{eqnarray}}
\newcommand{\eea}{\end{eqnarray}}
\newcommand{\beas}{\begin{eqnarray*}}
\newcommand{\eeas}{\end{eqnarray*}}

\newcommand{\bear}{\begin{array}{l}}
\newcommand{\eear}{\end{array}}

\newcommand{\bcf}{\begin{center}\begin{figure}}
\newcommand{\ecf}{\end{figure}\end{center}}

\newcommand{\bct}{\begin{center}\begin{table}}
\newcommand{\ect}{\end{table}\end{center}}

\newcommand{\ds}{\displaystyle}

\def\eq#1{(\ref{eq:#1})}

\def\Eqn#1{Equation~(\ref{eq:#1})}

\def\eqs#1#2{(\ref{eq:#1}) and~(\ref{eq:#2})}

\def\sec#1{section~\ref{sec:#1}}

\def\fig#1{figure~\ref{fig:#1}}

\newcommand{\Int}[1]{\int \!\! d^D \! #1 \,}

\newcommand{\MomInt}[2]{\int \!\! \frac{d^{#1} #2}{(2\pi)^{#1}} \, }

\newcommand{\der}[2]{\ensuremath{\frac{d #1}{d #2}}}
\newcommand{\pder}[2]{\ensuremath{\frac{\partial #1}{\partial #2}}}
\newcommand{\fder}[2]{\ensuremath{\frac{\delta #1}{\delta #2}}}
\newcommand{\Or}{\mathcal{O}}

\newcommand{\order}[1]{\Or ( #1 )}
\newcommand{\hf}{\frac{1}{2}}

\newcommand{\measure}[1]{\mathcal{D} #1 \, }

\newcommand{\DD}[1]{\delta^{(#1)}}

\newcommand{\SU}{\mathrm{SU}}

\def\dd{\dot{\Delta}}

\newcommand{\hS}{\hat{S}}
\newcommand{\hSR}{\hat{S}^{\mathrm{I}}}
\newcommand{\SigmaR}{\Sigma^{\mathrm{I}}}

\def\one{\hbox{1\kern-.8mm l}}

\newcommand{\flow}{\Lambda \partial_\Lambda}
\newcommand{\SR}{S^{\mathrm{I}}}
\newcommand{\SRv}[1]{S^{\mathrm{I}(#1)}}

\newcommand{\dual}{\mathcal{D}}
\newcommand{\dualv}[1]{\dual^{(#1)}}

\newcommand{\dualvm}[1]{\dual_m^{(#1)}}

\newcommand{\dualvb}[1]{\dual_\mathrm{bare}^{(#1)}}
\newcommand{\dopi}{\overline{\dual}}
\newcommand{\dopiv}[1]{\overline{\dual}^{(#1)}}

\newcommand{\dopivm}[1]{\overline{\dual}_m^{(#1)}}

\newcommand{\dep}{\tilde{\Delta}}

\newcounter{Diagrams}
\Alph{Diagrams}
\newtheorem{Diag}{}[Diagrams]

\newlength{\VertexWidth}

\newlength{\LabLength}

\newcommand{\arXiv}[2]{arXiv:{#1} [#2]}

\newcommand{\hepth}[1]{hep-th/#1}

\newcommand{\heplat}[1]{hep-lat/#1}

\begin{document}

\preprint{DIAS-STP-08-09}

\title{Triviality from the Exact Renormalization Group}

\author{Oliver J.~Rosten}
\email{orosten@stp.dias.ie}
\affiliation{Dublin Institute for Advanced Studies, 10 Burlington Road, Dublin 4, Ireland}

\begin{abstract}
	Using the exact renormalization group, it is shown that no physically
	acceptable non-trivial fixed points, with positive anomalous dimension,
	exist for (i) O$(N)$ scalar field theory 
	in four or more dimensions, (ii) non-compact, pure Abelian gauge theory in any dimension.
	It is then shown, for both theories in any dimension, that otherwise physically acceptable 
	non-trivial fixed points with negative anomalous dimension are non-unitary.
	In addition, a very simple demonstration is given, directly from the exact renormalization
	group, that should a critical
	fixed point exist for either theory in any dimension, then the $n$-point correlation
	functions exhibit the expected behaviour.
\end{abstract}


\maketitle

\section{Introduction}

The renormalizability of quantum field theories in the nonperturbative, Wilsonian sense, is
determined by the existence, or otherwise, of critical fixed points and the renormalized trajectories
emanating from them~\cite{Wilson,TRM-Elements}. As a consequence of this, low energy effective theories which na\"{\i}vely 
appear nonrenormalizable could in fact be descended from some
ultraviolet (UV) fixed point. This is the idea behind the asymptotic safety scenario~\cite{Weinberg-AS}.
In this paper, we will rule out such a scenario for 
(i) O$(N)$ scalar field theory\footnote{Non-linear sigma models are not considered.} 
in or above four dimensions,
(ii) non-compact\footnote{As opposed to the compact formulation~\cite{Polyakov-NC_1,Polyakov-NC_2}; see \sec{U1}.}, pure Abelian gauge theory in any dimension, by showing that
no physically acceptable non-trivial fixed points exist in either case. 
Consequently, triviality of these theories follows from the very well known fact that the 
respective Gaussian fixed points do not support non-trivial renormalized trajectories.\footnote{
There are claims that the Gaussian fixed point of scalar theory supports relevant, interacting directions in $D=4$~\cite{Halp+Huang} but, as convincingly argued by Morris~\cite{TRM-Comment,TRM-3D,TRM-Elements}, these directions cannot be used
to construct a renormalized trajectory.}

There are two criteria we use to determine the physical acceptability of a fixed point. The
first is `quasi-locality'~\cite{ym}: we demand that the action has a derivative
expansion. Anticipating that we will be working in Euclidean space, the second is 
that the theory makes sense as a unitary quantum field theory, upon continuation to Minkowski space.


The analysis of the existence, or otherwise, of non-trivial fixed points is split into two parts. First, we consider
the case where the fixed point anomalous dimension, $\eta_\star$, is greater than or equal
to zero. We will demonstrate that there are no quasi-local fixed points  for the theories mentioned above,
given the restriction on dimension for the scalar case.
In the case of negative anomalous dimension, our results in fact apply in all dimensions, even in the
case of O$(N)$ scalar field theory. It has been known for
a long time that exotic Gaussian fixed points without the standard $p^2$ kinetic piece---going instead
as $p^{2n}$ for $n$ an integer greater than one---have negative anomalous dimensions.
However, such fixed points can be excluded
from our considerations by the requirement that the theory be physically acceptable, since the
absence of the standard kinetic term leads to violation of unitarity. It is worth noting that, from a
condensed matter point of view, such a requirement
is, of course, an irrelevance. However, even in this context
these fixed points are still unimportant in O$(N)$ scalar field
theory. This is because, as shown
by Wegner~\cite{Wegner_CS}, they have an infinite number of relevant directions for $D\leq4$ and so,
for typical condensed matter systems of interest, it is scarcely possible to approach the critical point.

Nevertheless, this says nothing as to the possible existence of \emph{non-Gaussian} fixed points with
negative anomalous dimension. In this paper, we will not show that such fixed points,
which satisfy the requirement of having a quasi-local action, do not exist. Rather, it will be shown
that should such fixed points exist, then they necessarily violate unitarity.
It is well worth noting that, in the vicinity of
a nonperturbative fixed point, we cannot rule out a negative anomalous dimensions by the \emph{usual}
unitarity arguments (upon continuation to Minkowski space). Given field strength renormalization, $Z$,
these arguments relate the unitarity constraint $0\leq Z \leq 1$ to a positive anomalous dimension
via a perturbative calculation; but there is no reason to believe a perturbative calculation near to a nonperturbative fixed point (see~\cite{Etsoku} for an interesting discussion on negative anomalous dimensions).

Whilst the results obtained in this paper are unsurprising, they are either
complimentary to or stronger than results obtained elsewhere.
The first rigourous proof of triviality in scalar field theories (with field denoted by $\varphi$) was provided
by Aizenman~\cite{Aizenman}, who considered a lattice $\lambda \varphi^4$ model and showed that
no interacting continuum limit exists in $D>4$. This result was confirmed and extended by Fr\"{o}hlich~\cite{Frohlich-Trivial}, who also proved that for one or two-component
lattice $\lambda \varphi^4$  models in $D=4$, the only non-trivial continuum limit 
would have to be asymptotically free, contradicting perturbative expectations. 
For a review of these ideas, see~\cite{Callaway}.

The approach taken in this paper is different, both in implementation and philosophy (see \sec{formalism} for a description of the formalism). First
of all, everything is done directly in the continuum.
Secondly, we are freed from considering a particular model, such as $\lambda \varphi^4$, and
analysing whether a continuum limit exists. Rather, the Wilsonian view point is adopted that the bare action of a nonperturbatively renormalizable theory is something which should be \emph{solved} for, not something which should be put in by hand~\cite{TRM-Elements}. Along a renormalized trajectory,  the bare action is the  `perfect action'~\cite{perfect} in the vicinity of the ultraviolet (UV) fixed point of the theory, and so it is determined by the fixed point action---which must itself be solved for---and the integration constants associated with the relevant (including marginally relevant) directions. 

As to addressing the question of the existence of non-trivial fixed points
in O$(N)$ scalar field theories in $D=4$, there are a
number of studies which take, as a starting point, the same formalism
that is employed in this 
paper~\cite{H+H,TRM-Truncations}. However,  these earlier
works rely on a truncated derivative expansion of the effective action. Nevertheless, despite
the truncation, the space of possible interactions is infinite dimensional, so the power of this approach should not be underestimated. Within this approximation scheme, the space of truncated effective actions was scanned, numerically, with the result that the only fixed point found was the Gaussian one. It should be emphasised that this technique is very powerful
for uncovering non-trivial fixed points in $D<4$ and analysing their properties (see~\cite{B+B} for a comprehensive guide to the literature).

With regards to the triviality of non-compact, pure Abelian gauge theory, the strongest results to date
are those of Morris~\cite{TRM-U1}. Working in three dimensions (both with and without a Chern-Simons term) he considered truncated effective actions of the form $f(F_{\mu\nu})$, where $f$ is allowed to be any invariant function of its argument. It was then shown, analytically, that the only acceptable fixed-point is the Gaussian one.
In this paper, physically acceptable non-trivial fixed points   are ruled out
without any approximation, and in any dimension, representing a dramatic improvement on the current state of the art.

\section{Formalism}

The arguments in this paper crucially exploit scaling relations for the $n$-point correlation
functions.  One of the results of this paper is a very simple derivation, from first principles, that the
correlation functions at a fixed point do indeed take the form dictated by scale invariance.
To achieve this, we employ the exact renormalization group (ERG), which is basically the continuous version of Wilson's RG~\cite{Wilson}. A particularly simple form of the ERG, due to Polchinski~\cite{pol}, will
be described in the next section. After discussing how to compute correlation functions in this formalism,
a slightly different ERG equation will be introduced which is better suited to our
purposes.

\label{sec:formalism}
\subsection{The Polchinski Equation}

 Working in $D$-dimensional Euclidean space and
starting from some high energy scale, degrees of freedom are integrated out
down to a lower, effective scale denoted by $\Lambda$. During this process, the action evolves 
into the Wilsonian effective action, $S_\Lambda$,
such that it encodes the effects of the high momentum modes. 
The ERG equation determines how the Wilsonian effective action varies with $\Lambda$. A central ingredient is the ERG kernel, which provides the flow equation with its UV regularization. To this end, we introduce
the `effective propagator',
\be
	\Delta(p,\Lambda) = \frac{c(p^2/\Lambda^2)}{p^2},
\label{eq:Delta}
\ee
where $c(p^2/\Lambda^2)$ is a UV cutoff function which dies off sufficiently rapidly for $p^2/\Lambda^2 \rightarrow \infty$, and for which
\be
c(0) = 1.
\label{eq:c0}
\ee 
The position-space kernel, $\Delta(x,y)$, is given by the Fourier transform of~\eq{Delta}. Note that we shall use $p$ to denote both a four-vector and its modulus, with the meaning hopefully being clear from the context.

Whilst the choice~\eq{Delta} is the typical one, we will temporarily work with
\be
	\Delta_m(p,\Lambda) = \frac{c(p^2/\Lambda^2)}{p^2 + m^2(\mu)},
\label{eq:massive}
\ee
where $\mu$ is an arbitrary scale and $m^2(\mu)$ is a mass parameter independent of $\Lambda$.
This mass term is included to provide infrared (IR) regularization though, as will be seen, it can usually
be dispensed with. 

In what follows, we will work with a single component scalar field, corresponding to the O$(1)$
model (\ie\ there is a $\varphi \rightarrow -\varphi$ symmetry); generalization to the multi-component case is trivial [note that full O$(N)$ symmetry is assumed at putative fixed-points].
We now define the interaction part of the Wilsonian effective action, $\SR_\Lambda[\varphi]$, according to
\be
	S_\Lambda[\varphi] = \hf \varphi \cdot \Delta^{-1}_m \cdot \varphi + \SR_\Lambda[\varphi].
\label{eq:full}
\ee
As usual, the dot notation is shorthand for the following:
$A \cdot B \equiv \Int{x} A(x) B(x)$. Similarly, $\varphi \cdot \Delta^{-1}_m \cdot \varphi \equiv \Int{x}\Int{y} \varphi(x) \Delta^{-1}_m(x,y) \varphi(y) = \Int{p}/(2\pi)^{D} \varphi(p) \Delta^{-1}_m (p) \varphi(-p)$.
Henceforth, we will cease to  explicitly indicate the $\Lambda$ dependence of $\SR$, for brevity.

The starting point for our analysis is the form of the ERG equation introduced by 
Polchinski~\cite{pol}:
\be
	-\flow \SR = \hf \fder{\SR}{\varphi} \cdot \dd_m \cdot \fder{\SR}{\varphi} 
	- \hf \fder{}{\varphi} \cdot \dd_m \cdot \fder{\SR}{\varphi},
\label{eq:Pol-Flow}
\ee
where the $\Lambda$-derivative is performed at constant $\varphi$
and the ERG kernel, $\dd_m$, is given by the flow of the effective propagator:
\be
	\dd_m \equiv -\Lambda \der{\Delta_m}{\Lambda}.
\label{eq:dd}
\ee

In addition to the proofs pertaining to triviality, 
the techniques developed below allow a very simple
derivation of the expected form of the two-point correlation function at a critical fixed point,
directly from the ERG.

\subsection{Correlation Functions}

We now define the `dual action', $\dual_m[\varphi]$, according to
\be
	- \dual_m[\varphi] = \ln
	\left\{
		\exp
		\left(
			\hf \fder{}{\varphi} \cdot \Delta_m \cdot \fder{}{\varphi}
		\right) 
		e^{-\SR[\varphi]}
	\right\},
\label{eq:dual}
\ee
where the subscript $m$ is to remind us that we have utilized $\Delta_m$ in the construction.

Before describing what this object represents, we will compute its flow. It is easy to confirm, 
using~\eqs{Pol-Flow}{dd}, that
\be
	-\flow \dual_m[\varphi] = 0.
\label{eq:dual-flow}
\ee
Thus, the dual action is an invariant of the ERG. Just like the Wilsonian effective action,
we can expand the dual action in powers of the fields, thereby defining its vertices,
each of which are separately invariants of the ERG:
\begin{multline}
		\dual_m[\varphi] = \sum_{n} \frac{1}{n!} \MomInt{D}{p_1} \cdots \MomInt{D}{p_n}
		\dualvm{n}(p_1,\ldots,p_n)
	\\
	\varphi(p_1)\cdots \varphi(p_n) \hat{\delta}^{(D)}(p_1+\cdots+p_n),
\end{multline}
where $\hat{\delta}^{(D)}(p) \equiv (2\pi)^{D} \DD{D}(p)$.
At the two-point level, we define $\dualvm{2}(p) \equiv \dualvm{2}(p,-p)$.

As we will shortly demonstrate, the vertices of the dual action are essentially $n$-point connected correlation functions. Indeed, from this perspective it is clear why these vertices are ERG invariants: such objects, which incorporate all quantum fluctuations, must be independent of the effective scale, $\Lambda$, which is just an introduced as an intermediate step to facilitate the evaluation of the partition function. 

To arrive at this interpretation of the dual action, it is useful to introduce a diagrammatic representation, about which three very important points should be made. First, the diagrammatics utilize exact vertices of the Wilsonian effective action, no perturbative expansion of the vertices having been performed.
Secondly, the diagrammatic expansion will never be truncated.
Thirdly, the dual action exists entirely independently of its diagrammatic representation. Throughout this paper, we will perform various manipulations of the dual action using the diagrammatics. However, it should be emphasised that exactly the same results could be obtained directly from a power expansion of~\eq{dual}, together with a field expansion of the Wilsonian effective action.\footnote{It is worth pointing
out that truncated field expansions are known to suffer deficiencies when looking for fixed
points~\cite{TRM-Truncations}, though see~\cite{Aoki}. In this paper, no truncation is ever
performed, and it is assumed that the conclusions drawn are reliable.} The point is that, as usual, the diagrammatics provide an intuitive and transparent means of performing these manipulations; but the use of this tool is by no means a necessity.

From~\eq{dual}, the dual action comprises all connected diagrams built out of
vertices of the interaction part of the Wilsonian effective action and effective propagators
(it is the logarithm which, as usual, ensures connectedness). Indeed, this diagrammatic form has
been used for some time~\cite{mgiuc,evalues,univ,NonRenorm,Resum}, without realising that it was a 
representation of~\eq{dual}.
A selection of terms contributing to the two-point vertex of the dual action is shown in \fig{terms}.
\bcf[h]
	\[
	\dualvm{2} = \ensuremath{\begin{array}{c}\begin{picture}(0,0)%
\epsfig{file=pstex/ReducedWEA-2.pstex}%
\end{picture}%
\setlength{\unitlength}{3947sp}%
\begingroup\makeatletter\ifx\SetFigFont\undefined%
\gdef\SetFigFont#1#2#3#4#5{%
  \reset@font\fontsize{#1}{#2pt}%
  \fontfamily{#3}\fontseries{#4}\fontshape{#5}%
  \selectfont}%
\fi\endgroup%
\begin{picture}(358,579)(1629,-672)
\put(1730,-448){\makebox(0,0)[lb]{\smash{{\SetFigFont{11}{13.2}{\rmdefault}{\mddefault}{\updefault}{\color[rgb]{0,0,0}$\SR$}%
}}}}
\end{picture}%
 \end{array}} + \frac{1}{2} \ensuremath{\begin{array}{c}\begin{picture}(0,0)%
\epsfig{file=pstex/Padlock-2.pstex}%
\end{picture}%
\setlength{\unitlength}{3947sp}%
\begingroup\makeatletter\ifx\SetFigFont\undefined%
\gdef\SetFigFont#1#2#3#4#5{%
  \reset@font\fontsize{#1}{#2pt}%
  \fontfamily{#3}\fontseries{#4}\fontshape{#5}%
  \selectfont}%
\fi\endgroup%
\begin{picture}(418,565)(1606,-593)
\put(1727,-457){\makebox(0,0)[lb]{\smash{{\SetFigFont{11}{13.2}{\rmdefault}{\mddefault}{\updefault}{\color[rgb]{0,0,0}$\SR$}%
}}}}
\end{picture}%
 \end{array}} - \ensuremath{\begin{array}{c}\input{pstex/Dumbbell-2.pstex_t} \end{array}} 
	-\frac{1}{6} \ensuremath{\begin{array}{c}\input{pstex/TP-TL.pstex_t} \end{array}} + \cdots
	\]
\caption{The first few terms that contribute to $\dualv{2}$. Momentum arguments have been suppressed. Each of the lobes represents a vertex of the interaction part of the Wilsonian effective action.
}
\label{fig:terms}
\ecf

It is at this point we see that, by choosing a massive effective propagator, it has been ensured
that potentially IR divergent diagrams contributing to $\dual_m$ are individually regularized. These divergences 
have two sources. First, since $\dual_m$ contains one-particle reducible (1PR) terms, a massless
effective propagator would lead to strongly IR divergent diagrams in any dimension, as the external 
momenta go to zero. Furthermore,
depending on the dimensionality, one-particle irreducible (1PI) diagrams, such 
as the final diagram of \fig{terms} (which can also appear as a sub-diagram in 1PR terms), can also possess IR divergences. These latter
divergences do not occur exclusively for the external momenta going to zero, but can
also occur for other `special' momenta, whereby the external momenta entering a vertex sum
to zero.
However, since these divergences are also regularized by using a massive effective propagator, we will take our definition of IR divergences to include them.

Having made such a big deal of the IR regularization of the dual action we now argue that,
for most purposes, it is quite legitimate to deal with the vertices constructed from the massless
effective propagator. The first point to make is that the dual action plays a very different role from the
Wilsonian effective action. In particular, it does not appear as the weight in a partition function
where we are integrating over all field configurations. Indeed, if we think of it simply as the 
object which naturally collects together the vertices $\dualvm{n}$, then it does not necessarily
matter if the
\be
	\dualv{n}(p_1,\ldots,p_n) \equiv \lim_{m(\mu) \rightarrow 0} \dualvm{n}(p_1,\ldots,p_n)
\label{eq:no-m}
\ee 
have poles for certain values of their arguments.

With this in mind, consider computing connected $n$-point correlation functions from the \emph{bare} action.\footnote{The bare action is to be interpreted as in the introduction. For a more detailed discussion see~\cite{NonRenorm}, but these subtleties are of no real consequence here.} For $n>2$, the first contribution
comes from the $n$-point bare action vertex, connected to $n$ bare propagators, $\Delta_b$. Since this vertex is pulled down from $e^{-S_\mathrm{bare}}$, this contribution comes with a minus sign. Then we must include all other connected contributions with $n$ legs, built out of the bare action. Thus we find that
\[
	G(p_1,\ldots,p_n)
	=
	- \dualvb{n}(p_1,\ldots,p_n) \prod_{i=1}^{n} \Delta_b(p_i), \qquad n>2.
\]
But, we know that the dual action vertices do not depend on $\Lambda$, and so
\[
	G(p_1,\ldots,p_n)
	=
	- \dualvm{n}(p_1,\ldots,p_n) \prod_{i=1}^{n} \Delta_b(p_i), \qquad n>2.
\]
Consequently, $\dualvm{n}$ is directly related to the $n$-point connected correlation function
and so the limit  $m(\mu) \rightarrow 0$ makes perfect sense: any IR divergences that now appear
are just those we expect from the correlation functions.

Next, let us consider the two-point connected correlation function, $G(p)$, again computed
from the bare action. The first contribution to this is just the bare propagator. The full
contribution is
\be
	G(p) = \Delta_b(p) \left[1 -  \dualvm{2}(p)\Delta_b(p) \right]
\label{eq:ConnTP}
\ee
where, again, we have recognized that since the two-point dual action vertex is independent of
scale, we can evaluate it at the effective, rather than bare, scale.

We now introduce two objects which will play a central role in what follows. First, we define the
1PI components of the $\dualvm{n}$, denoted $\dopivm{n}$. 
The two-point object, $\dopivm{2}$, will play a special role.
As can be readily seen by considering the diagrammatic expression for $\dualvm{2}$ in more 
detail~\cite{mgiuc,NonRenorm}, $\dualvm{2}$ is built out
of $\dopivm{2}$ according to a geometric series:
\be
	\dualvm{2}(p) = \frac{\dopivm{2}(p)}{1 + \Delta_m(p) \dopivm{2}(p)}.
\label{eq:D2}
\ee
By inspection, this equation can be inverted:
\be
	\dopivm{2}(p) = \frac{\dualvm{2}(p)}{1 - \Delta_m(p) \dualvm{2}(p)}.
\label{eq:D2_invert}
\ee

The second key object is the dressed effective propagator,
$\dep_m$:
\be
	\dep_m(p) \equiv \frac{1}{\Delta^{-1}_m(p) + \dopivm{2}(p)}.
\label{eq:dep}
\ee
Substituting~\eq{D2_invert} into~\eq{dep} yields
\be
	\dep_m(p) = \Delta_m(p) \left[1 -  \dualvm{2}(p)\Delta_m(p) \right],
\label{eq:dep-alt}
\ee
and so we see that the dressed effective propagator is
a UV regularized version of $G(p)$. Consequently, all the objects $\dualvm{n}$ and $\dep_m$
makes sense in the limit $m(\mu) \rightarrow 0$, with any IR divergences having a physical
interpretation.

We now recognize that, as usual (see \eg~\cite{WeinbergI}), the physical mass of our theory is
defined by $\Delta^{-1}_m(0) + \dopivm{2}(0)$. Therefore, IR regularization---should we want it---really means that  $\Delta^{-1}_m(0) + \dopivm{2}(0) \neq 0$. In cases where we have performed the resummation~\eq{dep}, we now interpret any subscript $m$s to mean that we are constrained
to lie on a massive RG trajectory.

For the rest of this paper, we will send $m(\mu) \rightarrow 0$.
Nevertheless, there are certain circumstances where we should maintain the explicit IR regularization,
at least at intermediate stages. This should be done, for example, when inverting~\eq{dual}, to
recover the Wilsonian effective action:
\be
	- \SR[\varphi] = \ln
	\left[
		\exp
		\left(
			-\hf \fder{}{\varphi} \cdot \Delta_m \cdot \fder{}{\varphi}
		\right) 
		e^{-\dual_m[\varphi]}
	\right].
\label{eq:invert}
\ee
Incidentally, this relationship 
was proven diagrammatically in~\cite{NonRenorm} whereas here it follows trivially.

\subsection{Generalized ERGs}

Our aim now is to attempt to utilize the dual action to investigate the existence of fixed points.
Fixed point behaviour is most easily seen by rescaling to dimensionless variables, by
dividing all quantities by $\Lambda$ to the appropriate scaling dimension (by this it is meant, of course, the full scaling dimension, not just the canonical dimension). As it turns out, there is a well known subtlety related to
scaling
out the anomalous dimension from $\varphi$, so we will consider this rescaling first, in isolation.
Thus, we make the following transformation:
\be
	\varphi(x) \rightarrow \varphi(x)  \sqrt{Z},
\label{eq:rescale}
\ee
where $Z$ is the field strength renormalization, from which we define the anomalous dimension:
\be
	\eta \equiv \Lambda \der{\ln Z}{\Lambda}.
\label{eq:eta}
\ee
The problem with this transformation is that it produces an annoying factor of $1/Z$ on the
\rhs\ of the flow equation. However, we can remove this factor by utilizing the immense freedom
inherent in the ERG. General ERGs are defined according to~\cite{WegnerInv,TRM+JL}:
\be
\label{eq:blocked}
-\flow e^{-S[\varphi]} =  \int_x \fder{}{\varphi(x)} \left(\Psi_x[\varphi] e^{-S[\varphi]}\right).
\ee
The total derivative on
the \rhs\ ensures that the 
partition function $Z = \int \measure{\varphi} e^{-S}$
is invariant under the flow---a fundamental ingredient of any
Wilson-inspired
ERG equation. The functional, $\Psi$, parametrizes a
general Kadanoff blocking~\cite{Kadanoff} in the 
continuum and so there is considerable choice in its precise
from. We will focus on those blockings for which
\be
\label{eq:Psi}
	\Psi_x = \hf \Int{y} \dd^{\mathrm{new}}(x,y) \fder{\Sigma}{\varphi(y)},
\ee
with $\dd^{\mathrm{new}}$ not yet fixed to be given by either~\eq{Delta} or~\eq{massive} 
and 
\[
\Sigma \equiv S - 2\hS,
\]
where $\hS$ is the seed action~\cite{aprop,mgierg1,mgierg2,scalar2,qed,qcd}.
Whereas we solve the flow equation for the Wilsonian effective action,
the seed action serves as an input and, given our choice~\eq{Psi}
and  a choice of cutoff function,
parametrizes the remaining freedom in how modes are
integrated out along the flow. The only restrictions on the seed
action are that it leads to finite momentum integrals and that
it admits an all orders derivative expansion.  This latter property,  \aka\
`quasi-locality'~\cite{ym}, is a fundamental requirement of all ingredients of the ERG equation
and so covers $\dd$ and $S$, as well. Quasi-locality ensures that each ERG step is free
of IR divergences or, equivalently, that blocking is performed only over a local patch.

We now choose the new ERG kernel such that, \emph{after} performing the rescaling~\eq{rescale},
the flow equation reads:
\be
	\left(-\flow +\frac{\eta}{2} \varphi \cdot \fder{}{\varphi} \right) S 
	= \hf \fder{S}{\varphi} \cdot \dd \cdot \fder{\Sigma}{\varphi} 
	- \hf \fder{}{\varphi} \cdot \dd \cdot \fder{\Sigma}{\varphi}.
\label{eq:flow-new}
\ee
The next step in the analysis is to define the interaction part of the seed action,
analogously to~\eq{full}:
\be
\hS[\varphi] = \hf \varphi \cdot \Delta^{-1} \cdot \varphi + \hSR[\varphi].
\ee
Substituting this into~\eq{flow-new} yields, up to a discarded vacuum energy term,
\begin{multline}
	-\flow \SR + \frac{\eta}{2} \varphi \cdot \fder{S}{\varphi}
	=\hf \fder{\SR}{\varphi} \cdot \dd \cdot \fder{\SigmaR}{\varphi} 
\\
	- \hf \fder{}{\varphi} \cdot \dd \cdot \fder{\SigmaR}{\varphi}
	- \varphi \cdot \Delta^{-1} \cdot \dd \cdot \fder{\hSR}{\varphi}
\label{eq:reduced-flow}
\end{multline}
where $\SigmaR \equiv \SR - 2\hSR$. Notice that if we take $\hSR = 0$---as we are perfectly
at liberty to do---then~\eq{reduced-flow} is the rescaled version of Polchinski's equation, modulo
the fact that we have gotten rid of the annoying factor of $1/Z$ on the \rhs. This flow equation was
first considered by Ball et al.~\cite{Ball}. The more general version, with a non-zero $\hSR$, has been considered
in~\cite{scalar1,scalar2,NonRenorm}.

Given the new flow equation, we retain our definition for the dual action~\eq{dual},
despite the fact that the fields have been rescaled according to~\eq{rescale}.
Using the new flow equation we therefore have:
\begin{multline}
	-\left(\flow + \frac{\eta}{2} \varphi \cdot \fder{}{\varphi} \right) \dual[\varphi] 
	= -\frac{\eta}{2} \varphi \cdot \Delta^{-1} \cdot \varphi
\\
	+ 
	e^{\dual} \varphi \cdot \Delta^{-1} \cdot \dd \cdot
	\exp
	\left(
		\hf \fder{}{\varphi} \cdot \Delta \cdot \fder{}{\varphi}
	\right) 
	\fder{\hSR}{\varphi} e^{-\SR},
\label{eq:dualflow-Seed}
\end{multline}
where we recall~\eq{no-m}.
It is well worth noting that the seed action appears only in a single term, all other
occurrences having cancelled out. These cancellations were previously demonstrated
using elaborate (though increasing sophisticated) diagrammatics~\cite{scalar2,Primer,NonRenorm,mgiuc,qcd,mgierg1,qed};
now, however, they follow from a few lines of algebra!
This equation obviously simplifies to
\be
	-\left(\flow + \frac{\eta}{2} \varphi \cdot \fder{}{\varphi} \right) \dual[\varphi] 
	= -\frac{\eta}{2} \varphi \cdot \Delta^{-1} \cdot \varphi,
\label{eq:Rflow}
\ee
in the case that we choose $\hSR = 0$, as we now do. We will
comment on this choice further at the end of the paper.
Notice that the relative sign between the two terms on the \lhs\ of~\eq{Rflow} has flipped, compared to~\eq{reduced-flow}. 

To conveniently uncover fixed point solutions, we need to complete the rescalings
started with~\eq{rescale}. To this end, we define the `RG-time',
\be
	t \equiv \ln \mu/\Lambda,
\ee
and  also scale out
the various
canonical dimensions:
\be
	\varphi(x) \rightarrow \varphi(x) \Lambda^{(D-2)/2}, \qquad p_i \rightarrow p_i \Lambda.
\label{eq:Canonical}
\ee
In these units, fixed point solutions follow from the condition
\be
	\partial_t S_\star[\varphi] = 0
\label{eq:FP}
\ee
since, if all variables are measured in terms of $\Lambda$, independence of $\Lambda$ implies
scale independence. (Subscript $\star$s will be used to denote fixed-point quantities.) 

Unlike the rescaling~\eq{rescale}, the rescalings~\eq{Canonical} do not introduce 
further subtleties concerning the form of the flow equation
which now reads:
\begin{multline}
	\left(\partial_t + d_\varphi \varphi \cdot \fder{}{\varphi} + \Delta_\partial - D \right) \SR
	\\
	=
	\fder{\SR}{\varphi} \cdot c' \cdot \fder{\SR}{\varphi}
	-
	\fder{}{\varphi} \cdot c' \cdot \fder{\SR}{\varphi}
	-\frac{\eta}{2} \varphi \cdot \Delta^{-1} \cdot \varphi,
\label{eq:RescaledFlow}
\end{multline}
where $d_\varphi \equiv (D-2+\eta)/2$ is the scaling dimension of the field, a prime denotes a derivative \wrt\ the argument and
$\Delta_\partial$ is the `derivative counting operator'~\cite{WegnerInv,TRM-Deriv} (utterly unrelated to the effective propagator, $\Delta$):
\[
	\Delta_\partial \equiv D + \MomInt{D}{p} \varphi(p) p_\mu \pder{}{p_\mu}  \fder{}{\varphi(p)}.
\]
We can remove the leading $D$ from this expression if we specify that the $\partial /\partial p_\mu$ does not
strike the momentum conserving $\delta$-function associated with each vertex. 

\Eqn{Rflow} becomes:
\begin{multline}
	\left(\partial_t + \frac{D-2-\eta}{2} \varphi \cdot \fder{}{\varphi} + \Delta_\partial - D \right) 
	\dual[\varphi] 
\\
	= -\frac{\eta}{2} \varphi \cdot \Delta^{-1} \cdot \varphi.
\label{eq:dual-flowR}
\end{multline}

Given all the rescalings, which in particular mean that $\Delta(p) = c(p^2)/p^2$,
it follows from the definition of the dual action~\eq{dual}---with $m(\mu)=0$---that the fixed point condition~\eq{FP} implies
\be
	\partial_t \dualv{n}_\star = 0.
\label{eq:DualFP}
\ee

\section{Critical Fixed Points in Scalar Field Theory}

To analyse fixed points using the dual action formalism, let us start by
solving~\eq{dual-flowR} for $\dualv{2}$ at a fixed point:
\be
	-\frac{2+\eta_\star}{2} \dualv{2}_\star(p) + p^2 \der{\dualv{2}_\star(p)}{p^2} 
	= -\frac{\eta_\star}{2} \Delta^{-1}(p).
\ee
The solution to this
equation is
\be
	\dualv{2}_\star(p) =
	-p^{2(1+\eta_\star/2)}
	\left[
		\frac{1}{b(\eta_\star)} + \frac{\eta_\star}{2} \int dp^2 \frac{c^{-1}(p^2)}{p^{2(1+\eta_\star/2)}}
	\right] \!, 
\label{eq:D2-solution}
\ee
where $-1/b(\eta_\star)$ is the integration constant (assumed to be finite) 
and is a functional of the cutoff function. 
In the case where $\eta_\star \neq 0$,
$b$ is defined by the form of $\dualv{2}_\star(p)$ taken if we perform
the indefinite integral by Taylor expanding the cutoff function. For $\eta_\star = 0$, we make
a choice such that the leading behaviour in the first case coincides with the behaviour
in the second case, as $\eta_\star \rightarrow 0$.
Thus, for small
momentum, we have
\be
	\dualv{2}_\star(p) = 
	\left\{
	\begin{array}{ll}
	\ds
	-\frac{1}{b} p^{2(1+\eta_\star/2)} + \left(p^2 + \mbox{subleading}\right), & \eta_\star \neq 0,
	\\[2ex]
	\ds
	\left(1 - \frac{1}{b}\right)p^2, & \eta_\star = 0.
	\end{array}
	\right.
\label{eq:Soln_2}
\ee
Note that the subleading terms are cutoff dependent, not just with regards to their
prefactors, but also to their structure. For example, if $\eta_\star = 2$ and
$c'(0) \neq 0$, then the
subleading piece has a nonpolynomial component $p^4 \ln p^2$, but this is absent
altogether if $c'(0) = 0$. However, so long as $\eta_\star <2$, the subleading term in the brackets is always
subleading compared to $b p^{2(1+\eta_\star/2)}$ and
substituting~\eq{Soln_2} into~\eq{ConnTP} yields
\be
	G(p)
	\sim
	\frac{1}{p^{2(1-\eta_\star/2)}} \sim \dep_\star(p),
\label{eq:small_p}
\ee
for small $p$. This is precisely the behaviour that we expect at a critical fixed point. 
Whilst there are simple, general arguments as to why such behaviour
is expected (see \eg~\cite{Fisher-Rev}), I am unaware of a derivation
as simple as this, directly from the ERG (see \eg~\cite{Wegner_CS} for a different ERG
derivation of~\eq{small_p} at the critical point of some model).
For $\eta_\star \geq 2$, the leading behaviour of $G(p)$ in the small $p$ limit no longer describes critical behaviour and
is, indeed, cutoff dependent. We do not consider this further (though it is straightforward enough to adapt the following analysis to deal with this scenario).

Note that in the large $p$ limit we have
\be
	\lim_{p \rightarrow \infty} \dep(p) = \frac{f(p^2)}{p^2},
\label{eq:large-p}
\ee
where $f(p^2)$ is a monotonically decreasing function, related to the cutoff function, with $f(p^2) \geq 0$
for real $p$ [this is most easily seen by using a power law cutoff $c^{-1}(p^2) = 1 + p^{2r}$ in~\eq{D2-solution} and substituting the result into~\eq{dep-alt}]. It is worth pointing out that~\eq{large-p} is true irrespective of the sign of $\eta_\star$ and so negative anomalous dimensions cannot obviously be ruled out at non-trivial fixed points, based simply on the form of $\dep$.

Moving on to dual action vertices with more than two legs, the fixed point
equation is 
\begin{multline}
	\left(
		n \frac{D-2-\eta_\star}{2} + \sum_{i=1}^{n} p_i\cdot\partial_{p_i} - D
	\right)
\\
	\dualv{n>2}_\star(p_1,\ldots,p_n) = 0.
\label{eq:FPE_n}
\end{multline}
This has solution
\be
	\dualv{n>2}_\star(p_1,\ldots,p_n) =  P^{(n)}_r(p_1,\ldots,p_n),
\ee
with
$
	P^{(n)}_r(\xi p_1,\ldots,\xi p_n) = \xi^r P^{(n)}_r(p_1,\ldots,p_n)
$
and
\be
	r = D - n  \frac{D-2-\eta_\star}{2}.
\label{eq:n-cond}	
\ee

\subsection{Fixed Points with $\eta_\star \geq 0$}

Let us now focus on the case where $2 > \eta_\star \geq 0$. To this end,
we now analyse $\dopiv{2}_\star(p)$. First, we note from~\eqs{D2_invert}{Soln_2} that the leading
behaviour in the small $p$ limit is 
\be
	\dopiv{2}_\star(p) =
	\left\{
	\begin{array}{ll}
	\ds
	b p^{2(1-\eta_\star/2)} - p^2 +\ldots, &  2 > \eta_\star > 0
	\\[2ex]
	\ds
	(b-1)p^2 + \ldots, & \eta_\star = 0
	\end{array} 
	\right.
\label{eq:dbar2-Leading}
\ee
Secondly, we recognize that we can resum sets of loop diagrams
contributing to $\dopiv{2}(p)$ such that all internal lines become dressed, as indicated in
\fig{dressed-TP}. 
\bcf[h]
	\[
	\dopiv{2} = \ensuremath{\begin{array}{c}\begin{picture}(0,0)%
\epsfig{file=pstex/ReducedWEA-2.pstex}%
\end{picture}%
\setlength{\unitlength}{3947sp}%
\begingroup\makeatletter\ifx\SetFigFont\undefined%
\gdef\SetFigFont#1#2#3#4#5{%
  \reset@font\fontsize{#1}{#2pt}%
  \fontfamily{#3}\fontseries{#4}\fontshape{#5}%
  \selectfont}%
\fi\endgroup%
\begin{picture}(358,579)(1629,-672)
\put(1730,-448){\makebox(0,0)[lb]{\smash{{\SetFigFont{11}{13.2}{\rmdefault}{\mddefault}{\updefault}{\color[rgb]{0,0,0}$\SR$}%
}}}}
\end{picture}%
 \end{array}} + 
	\frac{1}{2} \ensuremath{\begin{array}{c}\begin{picture}(0,0)%
\epsfig{file=pstex/Padlock-2-dressed.pstex}%
\end{picture}%
\setlength{\unitlength}{3947sp}%
\begingroup\makeatletter\ifx\SetFigFont\undefined%
\gdef\SetFigFont#1#2#3#4#5{%
  \reset@font\fontsize{#1}{#2pt}%
  \fontfamily{#3}\fontseries{#4}\fontshape{#5}%
  \selectfont}%
\fi\endgroup%
\begin{picture}(418,580)(1606,-593)
\put(1727,-457){\makebox(0,0)[lb]{\smash{{\SetFigFont{11}{13.2}{\rmdefault}{\mddefault}{\updefault}{\color[rgb]{0,0,0}$\SR$}%
}}}}
\end{picture}%
 \end{array}} 
	-\frac{1}{6} \ensuremath{\begin{array}{c}\input{pstex/TP-TL-dressed.pstex_t} \end{array}} + \cdots
	\]
\caption{Resummation of diagrams contributing to $\dopiv{2}$: the thick lines represent
dressed effective propagators, \eq{dep}.}
\label{fig:dressed-TP}
\ecf
Now, consider the following scenarios:
\begin{enumerate}
	\item $D>4$, $\eta_\star \geq 0$,
	
	\item $D=4$ and $\eta_\star > 0$.
\end{enumerate}
Assuming that the Wilsonian effective action vertices are Taylor expandable for small momenta---this being one of our requirements for physical acceptability---it is apparent by power counting that
\bea
	 \dopiv{2}_\star(p) & = & \mathrm{const} + f_0(p),
\\
	\der{}{p^2} \dopiv{2}_\star(p) & = & \mathrm{const} + f_1(p),
\eea
where $\lim_{p\rightarrow 0} f_{0,1}(p) = 0$.
The second relationship follows from
considering diagrams like the third one in \fig{dressed-TP}. This diagram is the prototype for
diagrams whose first derivative \wrt\ $p^2$ will diverge for certain dimensions and/or values of
$\eta_\star$. In the IR, the leading term from this diagram looks like
\[
	\Int{k} \Int{l} \frac{1}{[k^2(l+p)^2 (l+k)^2]^{1-\eta_\star/2}}.
\]
Differentiating \wrt\ $p^2$ will increase the degree of IR divergence by two, but this is still not enough, given the above conditions on $D$ and $\eta_\star$, to render the term IR divergent as $p\rightarrow 0$.

More generally, we have the following power counting in the IR. Given $I$ internal lines
and $V$ vertices, there are $L = I -V +1$ loops. If we differentiate with respect to $p^2$ a total
of
P times, then the degree of IR divergence is
\[
	\mathbb{D} \geq D(I - V + 1) - 2(1-\eta_\star/2) I - 2P,
\]
where we understand $\mathbb{D} > 0$ to be IR safe. Now, since all two-point vertices have
been absorbed into the dressed effective propagators, and since we have only even-legged
vertices [which is ensured by our assumption of O$(N)$ symmetry at putative fixed-points], each vertex must have at least four legs.  Given that there are two external legs,
this implies that
\[
	I \geq 2V-1.
\]
Consequently [for $D\geq 2(1-\eta_\star/2)$], we have
\[
	\mathbb{D} \geq (D-4)V + (2V-1) \eta_\star +2(1- P).
\]
Given the restrictions that either $D>4$ and $\eta_\star \geq 0$ or $D=4$ and $\eta_\star>0$
we see that, both for $P=0$ and $P=1$, there are no IR divergent diagrams.
Acting on $\dopiv{2}(p)$ with further derivatives \wrt\ $p^2$, it may be that the limit $p\rightarrow 0$ now diverges,  but this does not concern us here. Rather, we simply note that we can
write
\be
	 \dopiv{2}_\star(p) \sim \mathrm{const} + \order{p^2} + \mathrm{subleading},
\label{eq:leading}
\ee
where the subleading terms are not necessarily polynomial in $p$. For a critical
fixed point, we must set the constant piece equal to zero. Comparing the resulting
equation with the top line of~\eq{dbar2-Leading}
we deduce the following: (i) in $D>4$, any critical fixed point with non-negative $\eta_\star$ must have precisely $\eta_\star = 0$ (ii) in $D=4$ there are no critical fixed points with $\eta_\star >0$ and so,
again, if there are to be any non-trivial fixed points with $\eta_\star \geq 0$, they must saturate the inequality.\footnote{Note that whilst an infinite number of diagrams are involved in the above arguments,
we expect the series to be (re)summable: no perturbative expansion of the vertices has been performed and we recall that the diagrammatics is just a convenient way of visualizing manipulations of the dual
action, which is related to the exact correlation functions.}

To rule out fixed points with $\eta_\star = 0$,
let us consider
the four-point vertex of the dual action. We start by expressing $\dualv{4}$ in terms of 1PI pieces:
\be
	\dualv{4}(p_1,p_2,p_3,p_4) = 
	\frac{\dopiv{4}(p_1,p_2,p_3,p_4)}{
		\prod_{i=1}^4\left[1 + \Delta(p_i) \dopiv{2}(p_i)\right]
	}.
\label{eq:D4}
\ee
(Since this formula is generally valid, we have dropped the $\star$.)
The resummation of the decorations of the legs into the denominator
makes it clear that, since we have just shown that $\dopiv{2}_\star(p) \sim p^2$,
any IR divergences of $\dualv{4}_\star$ must occur within $\dopiv{4}_\star$.
As with $\dopiv{2}$, we can resum classes of
loop diagrams contributing to $\dopiv{4}$ such that all internal lines
become dressed. A selection of the resummed diagrams contributing to
$\dualv{4}$ is shown in
 \fig{dressed-4pt}.
\bcf[h]
	\[
	\dualv{4} = \ensuremath{\begin{array}{c}\begin{picture}(0,0)%
\epsfig{file=pstex/ReducedWEA-4-dressed.pstex}%
\end{picture}%
\setlength{\unitlength}{3947sp}%
\begingroup\makeatletter\ifx\SetFigFont\undefined%
\gdef\SetFigFont#1#2#3#4#5{%
  \reset@font\fontsize{#1}{#2pt}%
  \fontfamily{#3}\fontseries{#4}\fontshape{#5}%
  \selectfont}%
\fi\endgroup%
\begin{picture}(379,526)(1621,-642)
\put(1730,-448){\makebox(0,0)[lb]{\smash{{\SetFigFont{11}{13.2}{\rmdefault}{\mddefault}{\updefault}{\color[rgb]{0,0,0}$\SR$}%
}}}}
\end{picture}%
 \end{array}} - \frac{1}{4} \ensuremath{\begin{array}{c}\input{pstex/4pt-oneloop-dressed.pstex_t} \end{array}}
	+ \cdots
	\]
\caption{Resummation of diagrams contributing to $\dualv{4}$. In the final diagram we implicitly sum over the independent permutations of the external legs. The thick external lines
denote decorated legs, as in~\eq{D4}.}
\label{fig:dressed-4pt}
\ecf

From~\eqs{FPE_n}{n-cond}, the fixed point solution for $\dualv{4}$ is
\be
	\dualv{4}_\star(p_1,p_2,p_3,p_4) =  P^{(4)}_r(p_1,p_2,p_3,p_4),
\ee
with
\be
	r = 4-D + 2\eta_\star.
\label{eq:4cond}	
\ee
First let us consider $D>4$. 

The crucial point is that, by the same logic that lead to~\eq{leading},
we have:
\be
	\dopiv{4}_\star(p_1,p_2,p_3,p_4) = c_4+ \mathrm{subleading},
\label{eq:leading-4}
\ee
where $c_4$ is a constant.
We can see this by looking at the second diagram on the \rhs\ of \fig{dressed-4pt} (modulo the dressings on the external legs), which is the prototype
for diagrams which possess IR divergences for certain values of $D$ and/or $\eta_\star$.
As a direct consequence of~\eq{leading-4}, it must be that $r \geq 0$.
But, given this condition and given $\eta_\star = 0$, \eq{4cond} has 
no solutions and so we conclude
that $\dualv{4}_\star = 0$. But, directly from this, it follows that $\dualv{6}_\star = 0$, also.
This is because in $D>4$ the only
contributions to $\dualv{6}_\star$ which could potentially violate the analogue of~\eq{leading-4}
comes from the diagram of \fig{D6-1PR}. But this term is built from 
a pair of $\dopiv{4}_\star$s, which we have just said vanish! Thus, 
repeating the same logic that lead to~\eq{4cond},
 we find that $\dualv{6}_\star = 0$. By induction, then, we have that $\dualv{n>2}_\star = 0$ and the only
 fixed point is the Gaussian one.
\bcf[h]
	\[
	\ensuremath{\begin{array}{c}\input{pstex/D4-dep-D4.pstex_t} \end{array}}
	\]
\caption{A potentially strongly IR divergent contribution to $\dualv{6}$.}
\label{fig:D6-1PR}
\ecf


For $D=4$, we could at this point conclude our analysis: after continuing to Minkowski
space, there is a theorem due to Pohlmeyer which implies that a scale invariant 
scalar field theory with vanishing anomalous dimension must be trivial~\cite{Pohlmeyer}. 
However, since we can demonstrate this directly within the current approach, with only a 
little extra effort, we will do so.
Consider the fixed point equation for $\dualv{n}_\star$:
returning to~\eq{FPE_n} there is now a solution
for the four-point vertex in $D=4$ with $\eta_\star =0$
which is potentially compatible with the structure of $\dualv{4}$:
\be
	\dualv{4}_\star(p_1,p_2,p_3,p_4) = c_4.
\ee
[Whilst contributions of the form \eg\ $p_1^2 / p_2^2$  are also solutions of~\eq{FPE_n}, there
is no way to generate such a strong IR divergence in $D=4$, given that $\eta_\star=0$.]
However, it turns out that $c_4$ must be zero. To see this, consider further building up contributions
to the second diagram in \fig{dressed-4pt}, as shown in \fig{4pt-dressedx2} (we drop the overall combinatoric factor).
\bcf[h]
	\[
	\ensuremath{\begin{array}{c}\input{pstex/Dbar4x2-depx2.pstex_t} \end{array}} + \hf \ensuremath{\begin{array}{c}\input{pstex/Dbar4x3-depx4.pstex_t} \end{array}} + \cdots
	\]
\caption{Further resummation of contributions to $\dualv{4}$.}
\label{fig:4pt-dressedx2}
\ecf

To justify this, let us start by considering the origin of the first term. Returning to the second
diagram of \fig{dressed-4pt}, imagine temporarily pulling the vertices off the pair of internal lines.
Each of these vertices has four legs: two external and two which, upon reattachment, are internal.
Now, rather than the pair of internal lines simply attaching at both ends to a single, undecorated vertex, 
we should also consider terms in which this pair of lines attaches to all possible structures. Summing over all such contributions, 
our
first thought might be that the pair of internal lines should attach at each end to a $\dualv{4}$,
as this is precisely the sum of all structures with a total of four legs. However, decorations
of the external legs have already been included (indicated by the thick external lines in \fig{dressed-4pt})
and the internal decorations have already been `used up' to dress the internal lines. Thus, our
second thought might be to attach the pair of internal lines at each end to a $\dopiv{4}$; indeed,
this yields precisely the first term in \fig{4pt-dressedx2} (up to the ignored overall constant). 

However, even this is not quite correct, as certain terms are wrongly counted. Let us return once
again to the second diagram of \fig{dressed-4pt} but this time imagine pulling off just the top vertex.
Let us now replace this vertex with a copy of the diagram we have just mutilated. From \fig{dressed-4pt}, 
this diagram comes with a factor of $-1/4$. However, we can attach it to the rest of the diagram via either
of its vertices, giving a net factor of $-1/2$. [The structure of this term is the same as that of
the second diagram in \fig{4pt-dressedx2}, but with the vertices
being just $\SRv{4}$s.] Now, were we to repeat this procedure by instead pulling off the
bottom vertex, we would generate an identical term. This symmetry
means that we should count the diagram only once.\footnote{An alternative way to see this is to
consider the combinatorics for building up such a term from scratch---see~\cite{NonRenorm}.} However, two copies are needed to build
up each of the $\dopiv{4}$s in the first diagram of \fig{4pt-dressedx2}. Thus, the first diagram
in \fig{4pt-dressedx2} overcounts the contribution discussed. To remedy this, we add the
next term in \fig{4pt-dressedx2} with the factor adjusted to precisely remove one of the
terms, of the structure just discussed, which is buried in the first diagram.

But we are still not done! For example, terms with three $\SRv{4}$s are not correctly counted.
To get the correct combinatorics for every contribution, we must add an infinite series of terms. 
The
good news is that these terms can be resummed. Indeed, we have met this problem before.
Since each vertex is joined to every other vertex by a pair of internal lines, imagine
representing this pair by a single line (say a red line). Similarly, combine the pair of (dressed) external legs on the topmost and bottom most vertices into a single red leg. Now, each
diagram is a string of `two-point' vertices. Moreover, we know that each of these `two-point'
vertices can be unpackaged into strings of various `two-point' 1PI diagrams each joined to the next by a red line.
Thus, modulo
the slightly different combinatorics, the job of resumming these terms is just the same as the
job of rewriting $\dopiv{2}$ in terms of $\dualv{2}$ \viz~\eq{D2_invert}. [This is the right
way around: each $\dualv{2}$ can be expanded out in terms of two point 1PI  diagrams---\ie\ the $\dopiv{2}$---each joined to the next by a single internal line.]

Taking the momenta carried by the top pair of legs to be $p$ and $q$, let us now define
\[
I(p,q)  \equiv \Int{k} \frac{\Delta^2(k) \Delta^2(k+p+q)}{\dep(k) \dep(k+p+q)}.
\]
Given~\eq{D4}, and since $\dualv{4}_\star = c_4$, we can resum the diagrams of \fig{4pt-dressedx2}
to give
\[
	\frac{c^2_4 I(p,q)}{1 -\hf c_4 I(p,q)}.
\]
This represents a momentum dependent contribution to $\dualv{4}_\star$ which cannot cancel against anything else. Therefore,
since $\dualv{4}_\star$ is a constant, we deduce that $c_4 = 0$ \ie\  $\dualv{4}_\star = 0$.

Now consider the
six-point dual action vertex. 
The solution to~\eq{FPE_n}, in $D=4$ with $\eta_\star=0$, requires that the six-point vertex $\sim 1/\mbox{mom}^2$. 
However,  such a divergence can only come from the diagram of \fig{D6-1PR}, which
again vanishes since $\dopiv{4}_\star = 0$. Consequently, we conclude that $\dopiv{6}_\star = 0$. Proceeding by induction, as before, we find
that the only acceptable critical fixed point solution in $D=4$, given the restriction that $\eta_\star$ is non-negative, is the Gaussian one!

Before moving on, it is instructive to examine the form of the Gaussian solution. In this case,
$\dopiv{2}_\star(p) = \SRv{2}_\star(p)$ and, using~\eq{full}, we find that
\[
	S_\star[\varphi] = \hf \varphi \cdot \frac{\Delta^{-1}(p)}{1 - (1-1/b) c(p^2)} \cdot \varphi,
\]
exactly in agreement with~\cite{TRM-Elements}.  The reason that there
is a line of equivalent Gaussian fixed points, parametrized by $b$, is due to the reparametrization invariance inherent in the ERG~\cite{WegnerInv,TRM-Elements,Wilson+Bell}.
Note that $b=1$ corresponds to canonical normalization of the kinetic term.

\subsection{Fixed Points with $\eta_\star < 0$}

The simplest fixed points with negative anomalous dimension to deal with are
the exotic Gaussian fixed points found by Wegner~\cite{Wegner_CS}. To recover
these, we set  $\dualv{n>2}_\star = 0$
and note that at a Gaussian fixed point we have $\dopiv{2}_\star(p) = \SRv{2}_\star(p)$.
Defining $\gamma_\star \equiv -\eta_\star > 0$ we can read off the leading behaviour
in the small momentum limit from~\eq{dbar2-Leading} by recognizing that this comes from the first
line, but with the relative importance of the first two terms interchanged:
\be
	\SRv{2}_\star(p) = - p^2 + b p^{2(1+\gamma_\star/2)}  + \ldots.
\ee
Now, the crucial point about the $-p^2$
term is that, as is apparent from~\eq{full}, it removes the standard kinetic term, $\hf  \varphi \cdot p^2 \cdot \varphi$, from the full Wilsonian effective action. Thus we find that, for small $p$,
\be
	S_\star^{(2)}(p) \sim  p^{2(1+\gamma_\star/2)},
\label{eq:WGFP}
\ee
where, to ensure locality, we must take $\gamma_\star/2$ to be an integer. 
This removal of the standard kinetic term is a generic feature of fixed points with
negative anomalous dimension, as we will now see.

Before embarking on an analysis of non-trivial fixed points with negative anomalous
dimension we note the following: if $\gamma_\star \geq D-2$ then diagrams such as the first one of 
\fig{dressed-TP} blow up independently of the external momenta (the external momentum does not
flow through the loop). Thus we only allow $D-2 > \gamma_\star >0$.

The vital property of fixed points with negative anomalous dimension, which we
will now exploit, is that
\be
	\lim_{p\rightarrow 0} \der{}{p^2} \dopiv{2}_\star(p) = -1,
\label{eq:minus_one}
\ee
and, since the \rhs\ is obviously independent of the shape of the cutoff function, the same
must be true of the \lhs. Note that
for fixed points with positive anomalous dimension, the \rhs\ of~\eq{minus_one}
instead diverges
and so the logic which we now use does not apply. This is a good job,
as otherwise we would rule out physically acceptable fixed points 
which we know to exist for $D<4$.

In what follows, it will be useful to define
\be
	z \equiv \lim_{p\rightarrow 0} \der{}{p^2} \SRv{2}(p).
\ee
Note that $z$ is just a number that we are free to chose, independently of the shape of the
cutoff function. Similarly to the above discussion, if $z > -1$, then the \emph{full} action
has a $p^2$ kinetic term with the right sign. In this case, $z$ is a free parameter
corresponding to the normalization of the field, with $z=0$ being canonical
normalization. The case where $z=-1$ 
includes Wegner's Gaussian fixed points but could, in principle, include
non-Gaussian fixed points. This could happen if all $n>2$-point vertices come with
sufficiently high powers of momenta on each of their legs. Either way, such fixed points
are ruled out by our requirement of unitarity, upon continuation to Minkowski space.
There is no need for us to consider $z<-1$, since the action then has a kinetic term of the
wrong sign. However, the following analysis anyway implies that non-trivial fixed points of this
type do not exist.

Let us denote the remaining contributions to the \lhs\ of~\eq{minus_one}
by $W$, so that we have
\be
	W = -z-1,
\ee
and focus now on the case where $z > -1$ is a finite number.

Again, we emphasise that $W$ is independent of the shape of the cutoff function.
Now, the only contribution to the cutoff function which is independent of its shape---\ie\ universal---is
$c(0) = 1$. Heuristically, then, we expect that any surviving contributions to $W$ come
from when the loop momenta are precisely equal to zero. To be more precise about this we
note that, since every contribution to $W$ contains at least one loop integral, we can write
\be
	W = \MomInt{D}{k} \frac{c(k^2)}{k^2} G[c](k^2).
\label{eq:W-struc}
\ee
Given that $\delta W/\delta c =0$ and noting that $W$ is a finite, nonzero number we have two options.
First, we could have that $G[c](k^2) = c^{-1}(k^2) g(k^2)$, where $\delta g(k^2) /\delta c = 0$ and
$\int_k g(k^2)/k^2 = -1-z$. However, if $g(k^2)$ is independent of the cutoff function, then this means that
$G[c](k^2)$ diverges strongly in the UV [so long as $c^{-1}(k^2)$ overwhelms $g(k^2)$ in this regime,
which we can always arrange to be the case]. But, as we will now argue, the strength of
this divergence is incompatible with the structure of $G$.

We can generate $G[c](k^2)$
by discarding the first diagram of \fig{dressed-TP}, cutting out one effective propagator from 
all subsequent terms, and taking the nascent legs to carry momenta $k$ and $-k$. Actually, this argument works just as well
if we redefine $G$ such that it is obtained by cutting out a dressed effective propagator.
The diagrams cut in this way are built out of interaction vertices and dressed effective propagators.
These latter components are UV regularized, so any UV divergence in $G[c](k^2)$ must occur
via the Wilsonian effective action vertices.
Now, the presence of the final term in the flow equation~\eq{RescaledFlow} tells us that
strong (\ie\ related to the inverse of the cutoff function) UV divergent behaviour does feed in to the flow equation at the two-point level. However, in the diagrams we are considering, two-point vertices only ever appear embedded in dressed effective propagators,
which we know are UV regularized; there is no problem here. Nevertheless, this divergent behaviour can feed into the higher point vertices via the first term on the \rhs\ of the flow equation. 
But, the structure of this term means that any $c^{-1}$ is  always accompanied by a $c'$, ameliorating the effect. Whilst it is true that $c' c^{-1}$ can diverge in the UV if $c$ dies off faster than an exponential, in such cases it does so only logarithmically, compared with $c$ itself. Thus, the diagrams
contributing to $G$ cannot reproduce the strength of the divergence $G[c](k^2) = c^{-1}(k^2) g(k^2)$ must have if $\delta g /\delta c = 0$.

Returning to the discussion under~\eq{W-struc}, the other option  is 
that $G[c](k^2)$ has net contributions only when both $k$, and also all momenta internal
to $G$, are zero. It is tempting to say that such contributions must have zero support but this is not true, \emph{a priori}, as
it is quite possible that individual terms contributing to $d \dopiv{2}_\star(p) / dp^2$ diverge as $p \rightarrow 0$.

However, inspired by the resummations shown in \fig{4pt-dressedx2}, let us resum the diagrams contributing to $\dopiv{2}_\star(p)$, yet further, as shown in \fig{2pt-dressedx2}.
\bcf[h]
	\begin{multline*}
	\dopiv{2} = \ensuremath{\begin{array}{c}\begin{picture}(0,0)%
\epsfig{file=pstex/ReducedWEA-2.pstex}%
\end{picture}%
\setlength{\unitlength}{3947sp}%
\begingroup\makeatletter\ifx\SetFigFont\undefined%
\gdef\SetFigFont#1#2#3#4#5{%
  \reset@font\fontsize{#1}{#2pt}%
  \fontfamily{#3}\fontseries{#4}\fontshape{#5}%
  \selectfont}%
\fi\endgroup%
\begin{picture}(358,579)(1629,-672)
\put(1730,-448){\makebox(0,0)[lb]{\smash{{\SetFigFont{11}{13.2}{\rmdefault}{\mddefault}{\updefault}{\color[rgb]{0,0,0}$\SR$}%
}}}}
\end{picture}%
 \end{array}} 
	\\
		+ \left( \frac{1}{2} \ensuremath{\begin{array}{c}\begin{picture}(0,0)%
\epsfig{file=pstex/Padlock-2-dressed.pstex}%
\end{picture}%
\setlength{\unitlength}{3947sp}%
\begingroup\makeatletter\ifx\SetFigFont\undefined%
\gdef\SetFigFont#1#2#3#4#5{%
  \reset@font\fontsize{#1}{#2pt}%
  \fontfamily{#3}\fontseries{#4}\fontshape{#5}%
  \selectfont}%
\fi\endgroup%
\begin{picture}(418,580)(1606,-593)
\put(1727,-457){\makebox(0,0)[lb]{\smash{{\SetFigFont{11}{13.2}{\rmdefault}{\mddefault}{\updefault}{\color[rgb]{0,0,0}$\SR$}%
}}}}
\end{picture}%
 \end{array}}
		-\frac{1}{4!} \ensuremath{\begin{array}{c}\input{pstex/6pt-4pt.pstex_t} \end{array}}
		+ \cdots 
	\right)
	-\frac{1}{6} \ensuremath{\begin{array}{c}\input{pstex/TP-TL-dressedx2.pstex_t} \end{array}} + \cdots
	\end{multline*}
\caption{Further resummation of diagrams contributing to $\dopiv{2}$. The brackets contain
terms in which both fields decorate the same vertex. The
second ellipsis represents diagrams built out of just $\dopiv{n>2}$ vertices.}
\label{fig:2pt-dressedx2}
\ecf

As before, there are corrections to the final term in \fig{2pt-dressedx2}, to avoid overcounting.
However, the crucial point is that all such contributions, as well as the rest of those included
in the second ellipsis, are built out of the 1PI dual action
vertices (with more than two legs) and dressed effective propagators. Now, we know from the above arguments that
we can analyse the various contributions to the \lhs\ of~\eq{minus_one} for small loop momenta.
From the scaling laws for the dual action vertices and the dressed effective propagator, the
small momentum behaviour of two-point diagrams built out of the $\dopiv{n>2}$ and $\dep$
is just that of $\dopiv{2}$ (this is easy to check). Thus, after differentiation \wrt\ $p^2$, these contributions go precisely as
\[
	p^{2(\gamma_\star/2)}
\]
at a fixed point and, therefore, vanish in the $p \rightarrow 0$ limit. 

Consequently, the only contributions to $W$ come from the diagrams enclosed by
the brackets in \fig{2pt-dressedx2}. Even after differentiation \wrt\ $p^2$, 
these are most certainly IR safe for $p \rightarrow 0$, since
the external momentum never flows around any of the loops and so the diagrams 
do have zero support for 
vanishing loop momenta. Therefore,
there are no fixed points with $W \neq 0$. Finally, then, the only fixed points 
with negative anomalous dimension are those for which $z=-1$ and these
correspond to non-unitary theories, upon continuation to Minkowski space.\footnote{
It is worth pointing out that, for this argument to work, it is crucial that $z$ can be chosen independently
of the cutoff, which follows as a consequence of the freedom to choose the normalization of the field. 
Without this freedom, we would expect $z$ to be a functional of $c$, in which case $W$
could also be a functional of $c$, with only the sum $W+z$ independent of the shape of the cutoff function.
}

\section{Non-compact, pure Abelian Gauge Theory}
\label{sec:U1}

Exactly the same methodology can be applied to demonstrate that
there are no physically acceptable non-trivial fixed points
in non-compact, pure Abelian gauge theory, but this time in any dimension.
In a lattice formulation, compactness refers to the gauge connection,
$A_\mu$, being valued on a circle, as opposed to the real line.
In the continuum limit, the compact case supports the existence
of  field configurations corresponding to monopoles.

We focus on just the non-compact case, for which
a manifestly gauge invariant flow equation is
provided by~\cite{ym1}:
\be
	-\flow S[A] = \hf \fder{S}{A} \cdot \dd \cdot \fder{\Sigma}{A} 
	- \hf \fder{}{A} \cdot \dd \cdot \fder{\Sigma}{A},
\label{eq:A-Flow}
\ee
where the dots now include a contraction of the Lorentz indices, wherever these indices are suppressed. Manifest gauge invariance follows since, for the Abelian Abelian symmetry, $\delta /\delta A$ is gauge invariant. Gauge invariance
of the functional derivatives also means that the flow equation can be regularized just as in the
scalar case, simply by introducing a cutoff function $c(p^2/\Lambda^2)$.

Seeing as we have not fixed the gauge, it is 
important that
the effective propagator is properly interpreted. The
point is that, first and foremost, the object $\dd$ in the flow equation should be thought of
as an ERG kernel, which provides UV regularization, ensuring that the flow equation is well
defined; and if the flow equation is well defined, then we are happy. In scalar field theory, it just so happens
that $\dd$ can be directly interpreted as the flow of a UV regularized propagator. In manifestly
gauge invariant ERGs, 
the interpretation must be different~\cite{qcd,qed}, since we cannot define a propagator in the usual way, having never fixed the gauge. However, the integrated ERG kernel plays a role which is
analogous to the usual propagator, albeit residing inside ERG diagrams, rather than Feynman diagrams. Indeed, this is the reason for the terminology `effective propagator'. Actually, in the current
case of \emph{pure} Abelian gauge theory, these considerations are essentially irrelevant, as
we will see below.

With these points in mind, we define
\be
	S[A] = \hf A_\mu \cdot \Delta^{-1}_{\mu\nu} \cdot A_\nu + \SR[A]
\label{eq:A-full}
\ee
and also
\be
	\Delta(p) \equiv \frac{c(p^2)}{p^2}.
\label{eq:A-prop}
\ee
The consequence of manifest gauge invariance is that $\Delta(p)$ is the inverse of
$\Delta^{-1}_{\mu\nu}(p)$ only inverse in the transverse space:
\be
	\Delta^{-1}_{\mu\nu}(p) \Delta(p) = \delta_{\mu\nu} - \frac{p_\mu p_\nu}{p^2}
	=\frac{\Box_{\mu\nu}(p)}{p^2},
\ee
where $\Box_{\mu\nu}(p) \equiv p^2\delta_{\mu\nu} - p_\mu p_\nu$.
However, in the pure Abelian case, the final term, `the gauge remainder'~\cite{aprop}, has no
effect since gauge invariance implies that all vertices satisfy
\be
	p_{\mu_i} S_{\mu_1\cdots \mu_i\cdots\mu_n}(p_1,\ldots,p_i,\ldots,p_n) = 0, \qquad \forall i.
\label{eq:WID}
\ee
Consequently, whenever a gauge remainder appears inside a diagram, the diagram is annihilated.

As before, we define the dual action according to
\be
	- \dual[A] = \ln
	\left\{
		\exp
		\left(
			\hf \fder{}{A} \cdot \Delta \cdot \fder{}{A}
		\right) 
		e^{-\SR[A]}
	\right\},
\label{eq:A-dual}
\ee
and, as before, the dual action is an ERG invariant. 
Similarly, we can introduce the 1PI contributions to the dual action
vertices, so long as we properly take account of the fact the the two-point
dual action vertex is transverse. Thus, defining $w$ according to
\be
	\dualv{2}_{\mu\nu}(p) \equiv w(p) \Box_{\mu\nu}(p), 
\label{eq:w}
\ee
and defining $\overline{w}$ as the corresponding 1PI piece, the same logic
that led to~\eq{D2} yields:
\be
	w(p) = \frac{\overline{w}(p)}{1 + c(p^2) \overline{w}(p)}.
\label{eq:wp}
\ee
Noting on account of~\eq{WID} that, when building up internal lines according to
$\Delta(p) \left[ \delta_{\mu \nu} - \Delta(p) \overline{w}(p) \Box_{\mu\nu}(p) + \cdots \right]$,
all $p_\mu p_\nu$ contributions can be effectively set to zero,
we define the dressed effective propagator according to:
\be
	\dep(p) \equiv \frac{1}{p^2 \left[ c^{-1}(p^2) + \overline{w}(p) \right]}.
\label{eq:A-DEP}
\ee

Exactly as in the scalar case, the next step is to rescale to dimensionless variables, having
changed flow equation to avoid annoying factors of $1/Z$ on the \rhs. Taking the seed action to possess only a kinetic term, the rescaled flow equation
for the dual action reads:
\begin{multline}
	\left(\partial_t + \frac{D-2-\eta}{2} A \cdot \fder{}{A} + \Delta_\partial - D \right) \dual [A] 
\\
	= -\frac{\eta}{2} A_\mu \cdot \Delta^{-1}_{\mu\nu} \cdot A_\nu.
\label{eq:A-dual-flowR}
\end{multline}
When computing $\dualv{2}_\star(p)$, we must remember to account for the $\Box_{\mu\nu}(p)$ buried in the \rhs. Employing~\eq{w}, we find that
\be
	w_\star(p) =
	p^{2\eta_\star/2}
	\left[
		b -\frac{\eta_\star}{2} \int dp^2 \frac{c^{-1}(p^2)}{p^{2(1+\eta_\star/2)}}
	\right].
\label{eq:wstar}
\ee
We now follow through the logic used in the scalar case but note that, on account of~\eq{WID}
we have that, for $\eta_\star \geq 0$,
\be
	\lim_{p\rightarrow 0} \overline{w}_\star(p) = \mathrm{const} 
\ee
\emph{in any dimension}. However, since for $2>\eta_\star\geq0$ equations~\eqs{wp}{wstar} imply that
\[
	\lim_{p\rightarrow 0} \overline{w}_\star(p) \sim p^{-2\eta_\star/2},
\]
we conclude that
 the anomalous dimension for a putative
critical fixed point with non-negative $\eta_\star$ is zero. Utilizing the Ward identity,
it is straightforward to show that this implies that $\dualv{4}_\star$ vanishes, but now in any dimension.
As before, we can proceed by induction to show that all $\dualv{>2}_\star$ vanish.
The analysis of fixed points with negative anomalous dimension exactly mirrors the scalar case.

Note that, in three dimensions, the analysis is unchanged by 
the presence of a Chern-Simons term (see~\cite{TRM-U1} for a discussion of the effects of such
a term in the context of the ERG).

\section{Conclusion}

The proofs presented in this paper
rely crucially on the scaling properties of correlation functions at putative non-trivial fixed points.
Whilst these scaling properties can be deduced, heuristically, on general grounds, we provide a simple
and direct derivation using the ERG. The ERG gives a precise meaning to the Wilsonian effective action, out of which we build the correlation functions, 
and provides a nonperturbative regularization scheme. The correlation functions, with a bare propagator amputated from each leg, are collected together into the `dual action',
given by~\eqs{dual}{A-dual} for scalar field theory and non-compact, pure Abelian gauge theory, respectively.

Nevertheless, without somehow fixing the anomalous dimension,
not much more can be said about these fixed points. The crucial realization is that, for the examples studied in this paper, the
behaviour of the two-point dual action vertex forces the anomalous dimension at a critical fixed point to be either zero
or negative. In the former case, we were able to show that 
the only critical fixed point theory of this type is trivial. In the latter case, whilst it might be that quasi-local critical fixed points exist, it is shown that they necessarily
violate unitarity; 
it is worth noting that the regularization scheme inherent to the ERG plays a crucial role in this.

An obvious set of questions to ask now is whether  these methods can be extended  to say something about: 
\begin{enumerate}
	\item critical fixed points in 
	scalar field theory for $D<4$;
	
	\item other theories.
\end{enumerate}
With regards to the first question it is certainly straightforward to identify the Wilson-Fisher
fixed point using an $\epsilon$-expansion (this has been explicitly checked). Whether
any new approximation scheme and/or some refinement of existing approximation schemes is
possible within the framework of this paper is left as an open question.

With regards to the second question, there are several major obstructions to applying the
methods here to more complicated theories. First of all, recall that if the seed action has
interactions then the flow of the dual action picks up an extra term on the \rhs\ [see~\eq{dualflow-Seed}].
In the cases considered in this paper, this term could be dropped, leaving a very simple equation
for the dual action. However, in any theory where the two-point function is related to any higher
point vertices by some symmetry---including, of course, QED and Yang-Mills theories---the seed action must possess interaction terms, in order that
this symmetry be preserved by the flow equation. In these situations, the dual action is no longer an
invariant of the ERG, and it is hard to see how to proceed. Indeed, if we use a direct
generalization of the dual action proposed in this paper for these theories, then
it  does not even satisfy the right symmetries. 
(Although in retrospect, it is apparent that what would be the $\order{p^2} \times 
\mathrm{nonpolynomial}$ part of the two-point dual action vertex was used to extract the $\beta$-function in both QED~\cite{qed,Resum}, $\SU(N)$ Yang-Mills~\cite{mgierg1,mgierg2,mgiuc}, and QCD~\cite{qcd}.)
Whether the formalism can be adapted to cope with these issues is
also left to the future.

However, there are at least a couple of places where one might be able to apply the techniques of this paper. First of all, it is reasonably straightforward to supersymmetrize the construction
and analyse the existence of non-trivial fixed points in theories of a scalar chiral superfield~\cite{Susy-Chiral}. 
Actually, using general arguments,
it has recently been proven that for an asymptotic safety scenario to exist
for the Wess-Zumino model, the associated non-trivial fixed point must have a negative
anomalous dimension~\cite{Safety}. This can be ruled out using the
methodology of this paper. More generally, an asymptotic safety scenario for
theories of a scalar chiral superfield---including those without a three-point contribution to the
superpotential---does not exist~\cite{Susy-Chiral}.

It might also be possible to investigate similar issues
in the context of non-commutative scalar field theory. Whilst, from an ERG perspective, 
we might worry about
how to deal with the non-locality inherent in this scenario, it turns out that one can transfer
to a matrix basis and construct a matrix version of the Polchinski equation. Indeed,
having added  a harmonic oscillator term to the action, Grosse and Wulkenhaar used precisely
this formalism  to demonstrate the perturbative renormalizability of the resulting model~\cite{G+W}. 
Trivially,
then, the dual action can be constructed; the challenge is to understand
what the criteria for physical acceptability of (the non-commutative analogue of)
fixed points are and what the procedure for constraining $\eta_\star$ becomes, in
the matrix base
(for a thoroughly
Wilsonian approach to noncommutative scalar field theory, see~\cite{RG+OJR}). This could be particularly interesting in light of the
claim that there does indeed exist a non-trivial fixed point in the Grosse and Wulkenhaar
model~\cite{Vanishing}.

\begin{acknowledgments}
I acknowledge IRCSET for financial support. I would like to thank Denjoe O'Connor
for some extremely valuable discussions and also Brian Dolan, Andreas J\"{u}ttner, Peter Kopietz, Werner Nahm, Hugh Osborn, Roberto Percacci and Christian S\"{a}mann. Thanks must also go
to an anonymous referee for several insightful comments.
\end{acknowledgments}

\bibliography{Invariants9}

\end{document}